\documentclass[manuscript]{aastex}
\usepackage{amssymb}
\renewcommand{\b}[1]{\mbox{\boldmath $#1$}}
\begin{document}
\title{OPTICAL DEPTH OF THE COSMIC MICROWAVE BACKGROUND DUE TO SCATTERING AND ABSORPTION}
\author{V. Krishan}
\affil{Indian Institute of Astrophysics, Bangalore-560034, India\\
and\\ Raman Research Institute, Bangalore-560080, India}
\email{vinod@iiap.res.in}
\date{}
\begin{abstract}
 It is shown that, in addition to the Thomson scattering, the absorption due to the electron-electron, electron-ion and the electron -atom collisions in a partially ionized cosmic plasma would also contribute to the optical depth of the cosmic microwave background (CMB).  The absorption depth depends on the plasma temperature and frequency of the CMB radiation. The absorption effects are prominent at the low frequency part of the CMB spectrum. These effects when included in the interpretation of the CMB spectrum  may necessitate a revised view of the ioniziation of the universe.
\end{abstract}
\textit{Subject Headings}: cosmic microwave background � cosmology � intergalactic medium-absorption

\section{Introduction }
The Thomson scattering is considered to be the major source of the optical depth of the cosmic microwave background. The estimates of the electron density or the ionization fraction , thus, undergo a continual revision with the availability of the new data which then impacts the scenarios of the time and place of the sources of ionization in the universe (Shull and  Venkatesan 2008 and references therein). Here the absorption of the cosmic microwave background in a partially ionized cosmic plasma due to the electron-electron, electron-ion and electron-atom collisons is presented as an additional source of the optical depth. The absorption cross-section and the ensuing optical depth are calculated in the next section. Using the recent model values of the various cosmological quantities obtained from the WMAP-5 measurements, as listed by Shull and Venkatesan (2008), the total optical depth due to electron scattering and the absorption is estimated for different redshift values. The paper ends with a brief conclusion.
\section{Collisional Absorption }
Consider the propagation of electromagnetic radiation through a partially ionized plasma consisting of electrons, protons and hydrogen atoms. The elctrons undergo oscillations at the radiation frequency and collsions with the other electrons, the protons and the hydrogen atoms. The oscillatory energy of the electrons is dissipated due to the collisions and the electromagnetic radiation suffers damping. A simple analysis of this process can be carried out by writing the equation of motion of an electron in the electric field of the electromagnetic radiation, neglecting the magnetic force in the nonrelativistic limit, as:
\begin{eqnarray}
m_{\rm e}\frac{\partial {\b V}_{\rm e}}{\partial t}=-e {\b E}~.
\end{eqnarray}
 Assuming a plane-wave form $\b E=\b E_0 \exp(-i\omega t)$  the oscillatory velocity of an electron is found to be $ \| V_{\rm e}\|=\frac{e E_0}{ m_{\rm e}\omega}$. The frictional force on an electron is $\b F_{vis}=-\nu m_{\rm e}\b V_{\rm e}$ where  $\nu=\nu_{ee}+\nu_{ei}+\nu_{eH}$ is the sum of the electron-electron, electron-ion and the
electron-atom collision frequencies. The power dissipated is then given by $ P=-\b V_{\rm e}\cdot \b F_{vis}= \nu m_{\rm e} V_{\rm e}^2= \nu\frac{e^2E_0^2}{m_{\rm e}\omega^2}$. The absorption cross-section can be found from $\sigma_a=\frac{P}{S}$ ( Rybicki and Lightman 1979 ) to be
\begin{eqnarray}
\sigma_a=\frac{4\pi c}{\omega}\frac{e^2}{m_{\rm e}c^2}\frac{\nu}{\omega}~,
\end{eqnarray}
where $S=\frac{c E_0^2}{4\pi}$ is the Poynting flux. We define the ratio $R$ of the absorption $\sigma_a$ and the Thomson cross-section $\sigma_T$ as :

\begin{eqnarray}
R= \frac{\sigma_a}{\sigma_T} = R_{ee}+ R_{ei}+R_{eH}=\frac{3 c}{2\omega}\left(\frac{e^2}{m_{\rm e}c^2}\right)^{-1}\frac{\left(\nu_{ee}+\nu_{ei}+\nu_{eH}\right)}{\omega}~
\end{eqnarray}
The collision frequencies are given by ( Khodachenko et al. 2004 )
\begin{eqnarray}
\nu_{eH}=10^{-15}n_H \left(\frac{8K_B T}{\pi m_{eH}}\right)^{1/2},\quad \nu_{ei}=5.89\times 10^{-24}n_i\Lambda\left(K_B T\right)^{-3/2}, \quad \nu_{ee}\approx 60 \nu_{ei}~
\end{eqnarray}

where $n_H$ is the density of the hydrogen atoms, $T$ is the plasma temperature, $m_{eH}=\frac{m_e m_H}{m_e+m_H}\approx m_e$ is the effective mass of the colliding particles, $n_i=n_e$ is the proton density equal to the electron density and  $\Lambda\approx 10$ is the Coulomb logarithm. Exhibiting explicitly the redshift dependence of the parameters, the total optical depth $\tau$, due to the Thomson scattering, $\tau_T$, and the absorption, of the CMB can be written as ( Shull and Venkatesan 2008):
\begin{eqnarray}
\tau=\tau_T+ \tau_{ee}+\tau_{ei}+\tau_{eH}=\int_{z_1}^{z_2}n_e{\left(z\right)}\sigma_T\left(1+R_{ee}\left(z\right)+R_{ei}\left(z\right)+R_{eH}\left(z\right)\right) \left(\frac{c}{H\left(z\right)\left(1+z\right)} \right)dz~
\end{eqnarray}
where $\tau_{ee}$, $\tau_{ei}$ and $\tau_{eH}$ are respectively the optical depths due to absorption by electron-electron, electron-ion and electron- atom collisions.
We write $\omega=2\pi\nu_{09}\times 10^9 \left(1+z\right) s^{-1}$, $T=T_{0r}\left(1+z\right)$ for high redshifts and $T=T_{0r}\left(1+z\right)^2$ for low redshifts. From the model used by Shull and Vekatesan (2008), $H(z)=H_0[\Omega_m(1+z)^3 +\Omega_{\Lambda}]^{1/2}$, $h = (H_0/100)$ km $s^{-1} Mpc^{-1}$, the densities : $n_H =[(1-Y)\Omega_b\rho_{cr}/m_H](1 + z)^3,
n_{He} = yn_H$ , and $n_e=x_e n_H(1 + y)$, if helium is singly ionized. A primordial helium mass fraction, $Y = 0.2477 \pm 0.0029$
(Peimbert et al. 2007) was assumed. The helium fraction by number is $y= (Y/4)/(1 - Y )\approx 0.0823$. The critical density is $\rho_{cr} = (1.8785 \times 10^{-29}h^2 $ g $cm^{-3})$. \

The absorption optical depth can also be derived by writing the refractive index $N$ of the partially ionized plasma as:
\begin{eqnarray}
N^2= \frac{k^2 c^2}{\omega^2}=1-\frac{\omega_{ep}^2}{\omega\left(\omega+i\nu\right)}, \quad \omega_{ep}^2=\frac{4\pi e^2 n_e}{m_e}~
\end{eqnarray}
where $\omega_{ep}$ is electron plasma frequency and $k$ is the complex wavevector. The imaginary part of $k$ gives the absorption per unit length which when integrated over the path length gives the optical depth.

\section{AT LOW AND INTERMEDIATE REDSHIFTS }

 For $T=T_{0r}\left(1+z\right)^2$ at low redshifts, the collision frequencies become
\begin{eqnarray}
\nu_{eH}=1.2\times10^{-14} \left(1+z\right)^{4}\Omega_b h^2 ,\quad \nu_{ei}= 7.85\times 10^{-5}x_e \Omega_b h^2, \quad \nu_{ee}\approx 60 \nu_{ei}~
\end{eqnarray}
The ratio $R$ of the absorption and the scattering cross-sections at low redshifts is found to be:
\begin{eqnarray}
R=\left[19.15 x_e \left(1+z\right)^{-2}+0.5\times 10^{-10}\left(1+z\right)^{2}\right]\frac{\Omega_b h^2}{\nu_{09}^2}~
\end{eqnarray}
The total optical depth at low redshifts, for $ \Omega_m \left(1+z\right)^{3}>> \Omega_{\Lambda}$ which is satisfied for $z > (\frac{\Omega_{\Lambda}}{\Omega_m})^{1/3}-1\approx 0.38$, can be estimated by integrating Eq.(5) from  say a redshift $z=1$ to $z_r$ the epoch of reionization. The optical depth is found to be: 
\begin{eqnarray}
\tau=0.0023\left[\left(\left(1+z_r\right)^{3/2}-2^{3/2}\right)+ \right.\nonumber\\
\left.57.45 \frac{<x_e>}{\nu_{09}^2}\left(-\left(1+z_r\right)^{-1/2}+2^{-1/2}\right)+ \right.\nonumber\\
\left.\frac{0.2\times 10^{-10}}{\nu_{09}^2}\left(\left(1+z_r\right)^{7/2}-2^{7/2}\right)\right]~
\end{eqnarray}
The first observation is that the absorption optical depth, unlike the Thomson optical depth is a function of the CMB  radiation frequency and is inversely proportional to the square of the frequency. At $z_r=7$, near the so called reionization redshift, the absorption depth due to electron - electron and electron-ion scattering becomes equal to the Thomson depth at the CMB radiation frequency $\nu_0= 1.73$ GHz. for ionization fraction $x_e\approx 1$. The electron-atom collisional process becomes insignificant.\\
The contribution of the absorption processes can be really appreciated by subtracting the frequency independent Thomson part from the total optical depth viz
\begin{eqnarray}
\tau-\tau_T=0.0023\left[\right.\nonumber\\
\left.57.45 \frac{<x_e>}{\nu_{09}^2}\left(-\left(1+z_r\right)^{-1/2}+2^{-1/2}\right)+ \right.\nonumber\\
\left.\frac{0.2\times 10^{-10}}{\nu_{09}^2}\left(\left(1+z_r\right)^{7/2}-2^{7/2}\right)\right]~
\end{eqnarray}
We find that during the so called dark age, $z\approx 100$ and the ionization fraction $x_e\approx 10^{-5}$, the electron-atom absorption becomes comparable to the sum of the electron-electron and electron-ion absorptions.
\section{AT HIGH REDSHIFTS }
For high redshifts with $T=T_{0r}(1+z)$ the collision frequencies become
\begin{eqnarray}
\nu_{eH}=1.2\times10^{-14} \left(1+z\right)^{7/2}\Omega_b h^2 ,\quad \nu_{ei}= 7.85\times 10^{-5}x_e  \left(1+z\right)^{3/2}\Omega_b h^2~
\end{eqnarray}
The ratio $R$ of the absorption and the scattering cross-sections at high redshifts is found to be:
\begin{eqnarray}
R=\left[19.15 x_e \left(1+z\right)^{-1/2}+0.5\times 10^{-10}\left(1+z\right)^{3/2}\right]\frac{\Omega_b h^2}{\nu_{09}^2}~
\end{eqnarray}
The total optical depth for large redshifts can be estimated by integrating Eq.(5) from a redshift $z_r$ to $z_f$ along with the condition $ \Omega_m \left(1+z\right)^{3}>> \Omega_{\Lambda}$. Assuming an averaged ionization fraction $<x_e>$ for a specified range of redshifts , the optical depth is found to be: 
\begin{eqnarray}
\tau= 2.12\times 10^{-3}<x_e> \left[\left(1+z_f\right)^{3/2}-\left(1+z_r\right)^{3/2}+\right.\nonumber\\
\left.0.65 \frac{ <x_e>}{\nu_{09}^2}\left(z_f-z_r\right)+\right. \nonumber\\
\left. \frac{0.45\times 10^{-12}}{\nu_{09}^2}\left(\left(1+z_f\right)^{3}-\left(1+z_r\right)^{3}\right)\right]~
\end{eqnarray}
For $z_f >> z_r$, we see that at $z_f=800, <x_e>\approx 10^{-2.1}$, (Shull and Venkatesan 2008), the total optical depth  becomes :
\begin{eqnarray}
\tau= 2.12\times 10^{-3}<x_e> \left(2.2\times 10^4+ \frac{4.26}{\nu_{09}^2}\right)~
\end{eqnarray}
where the first term is due to the Thomson scattering and the second due to the electron- electron and the electron-ion absorption processes. The electron-atom collisional process becomes insignificant. Thus the absorption processes contribute $10$ percent of the Thomson optical depth at a frequency $\nu_{09}=0.044$ or $\nu_0\approx 44$ MHz. At $z_f=1000, x_e=10^{-1.1}$, the optical depth due to the absorption processes becomes $10$ percent at $\nu_0\approx 130$ MHz and  equal to the Thomson optical depth at a frequency $\nu_0\approx 41$ MHz. Again we find that the optical depth $(\tau_{ee}+\tau_{ei})$ becomes comparable to the optical depth $\tau_{eH}$ near the recombination era at $z_f\approx 1000, <x_e>\approx 7\times 10^{-7}$. 
\section{CONCLUSION }
It is found that the electron- electron, electron-ion and the electron- atom collisional processes contribute significantly to the optical depth of the CMB in addition to the Thomson scattering. The absorption processes are temperature and  frequency dependent  and contribute significantly at frequencies in the range of tens to hundreds of MHz. Since concerted efforts are planned worldwide to study the low frequency part of the CMB spectrum, we believe the inclusion of the absorption processes discussed here becomes essential while interpreting the low frequency CMB spectrum.
\section{Acknowledgments}The author thanks Prof. Ravi Subrahmanyan for suggesting this problem during a discussion.
The author acknowledges gratefully the support received from Dr. B. A. Varghese for his help in the preparation of this manuscript.

\end{document}